# Prediction Challenge: Simulating Rydberg Photoexcited Cyclobutanone with Surface Hopping Dynamics based on Different Electronic Structure Methods

Saikat Mukherjee,[1]* Rafael S. Mattos,[2] Josene M. Toldo,[3] Hans Lischka,[4] Mario Barbatti[5,6]*

[1]*Aix Marseille University, CNRS, ICR, Marseille 13397, France; ORCID: 0000-0002-0025-4735*
[2]*Aix Marseille University, CNRS, ICR, Marseille 13397, France; ORCID: 0000-0003-0215-7100*
[3]*Aix Marseille University, CNRS, ICR, Marseille 13397, France; ORCID: 0000-0002-8969-6635*
[4]*Department of Chemistry and Biochemistry, Texas Tech University, Lubbock, TX 79409-1061 USA; ORCID: 0000-0002-5656-3975*
[5]*Aix Marseille University, CNRS, ICR, Marseille 13397, France; ORCID: 0000-0001-9336-6607*
[6]*Institut Universitaire de France, Paris 75231, France*

*\* Corresponding author: saikat.mukherjee@univ-amu.fr, mario.barbatti@univ-amu.fr*

This research examines the nonadiabatic dynamics of cyclobutanone after excitation into the $n$-3s Rydberg $S_2$ state. It stems from our contribution to the Special Topic of the Journal of Chemical Physics to test the predictive capability of computational chemistry against unseen experimental data. Decoherence-corrected fewest-switches surface hopping (DC-FSSH) was used to simulate nonadiabatic dynamics with full and approximated nonadiabatic couplings. Several simulation sets were computed with different electronic structure methods, including a multiconfigurational wavefunction (MCSCF) specially built to describe dissociative channels, multireference semiempirical approach, time-dependent density functional theory, algebraic diagrammatic construction, and coupled cluster. MCSCF dynamics predicts a slow deactivation of the $S_2$ state (10 ps), followed by an ultrafast population transfer from $S_1$ to $S_0$ (<100 fs). CO elimination (C3 channel) dominates $C_2H_4$ formation (C2 channel). These findings radically differ from the other methods, which predicted $S_2$ lifetimes 10 to 250 times shorter and C2 channel predominance. These results suggest that routine electronic structure methods may hold low predictive power for the outcome of nonadiabatic dynamics.

## 1  Introduction

Thanks to commendable scientific efforts globally over the last few decades, a handful of theoretical methods[1–5] have emerged to simulate excited-state photoinduced dynamics. These developments span from excited-states electronic structure theory to nonadiabatic dynamics methodologies.[6] However, a lack of comprehensive and uniform studies comparing the performance of these methods with recent experimental findings leaves the question of which method excels still unanswered. To address this very issue, The Journal of Chemical Physics proposed a double-blind prediction challenge to forecast the outcomes of an upcoming experiment performed using ultrafast electron diffraction,[7] where a gas sample of cyclobutanone (CB) molecule will be irradiated with a 200 nm light source targeting the Rydberg excited state ($n \rightarrow 3s$) and electron diffraction images will be collected during the process in variable step sizes.

This article stems from our participation in this prediction challenge, offering theoretical insights and simulated electron diffraction spectra of the photochemical processes of CB in the gas phase. Our simulations employed decoherence-corrected fewest-switch surface hopping (DC-FSSH)[1] combined with different routine electronic structure methods, including MCSCF, MRCI/ODM3, TDDFT, ADC(2), and CC2 (acronyms are defined in Computational Methods).

We worked with an array of methods to gauge their quality performance. All approaches are well established, have low computational costs, and, except for MCSCF, are of straightforward usage. Each approach was expected to offer specific advantages and technical difficulties. MCSCF and MRCI/ODM3 should suffer from instabilities caused by changes in the active space orbital composition during dynamics. MCSCF mostly neglects dynamic electron correlation. MRCI/ODM3 should poorly describe diffuse states[8] like $S_2$ in cyclobutanone. The single-reference character of TDDFT, CC2, and ADC(2) may cause trouble during dissociation. They are also not expected to work well near $S_1/S_0$ crossings.[1] CC2, as a non-Hermitian method, should not converge when $S_2$ and $S_1$ are degenerated.[9] ADC(2) is well known for incorrectly describing the excited state of carbonyl compounds.[10] Thus, given all these limitations in the excited state calculations, it is clear that the actual provocation of this prediction challenge is not so much about the choice of the nonadiabatic dynamics approach but about finding an electronic structure method inexpensive enough to be used for dynamics and still deliver robust results for states with strongly different diabatic characters.[11,12]

Cyclobutanone mesmerizes chemists. It offers unique and exciting photochemical behavior compared to other noncyclic and cyclic ketones of larger ring sizes, mainly due to its ring strain. Upon $S_1$ photoexcitation, CB undergoes the Norrish type-I reaction with the cleavage of the C-C bond in the alpha to the carbonyl position ($\alpha$−cleavage) in the excited state and subsequent multiple bond breaking and formation via diradical intermediates.[13–15]

From the experimental absorption UV spectra,[14,16,17] we learn that the $S_1$ state is a symmetry-forbidden $n \rightarrow \pi^\star$ transition ranging from 330 to 240 nm, with its maximum around 280 nm. The $S_2$ state is a Rydberg excitation $n \rightarrow 3s$ spanning from 206 to 182 nm. Photoexcited CB in the gas phase can follow two main photodissociation channels: the C2 and the C3 pathways. The C2 pathway forms ethylene ($C_2H_4$) and ketene ($CH_2CO$), as depicted in Figure 1. In contrast, the C3 pathway forms cyclopropane (c-

$C_3H_6$) and carbon monoxide (CO). The branching ratio of C3 and C2 products depends sensitively on the excitation wavelength. The C3:C2 product ratio was recorded as 0.8 at 248 nm, 0.4 at 313 nm, 2 at 326 nm, and 7 at 344 nm in the gas phase.[18–20]

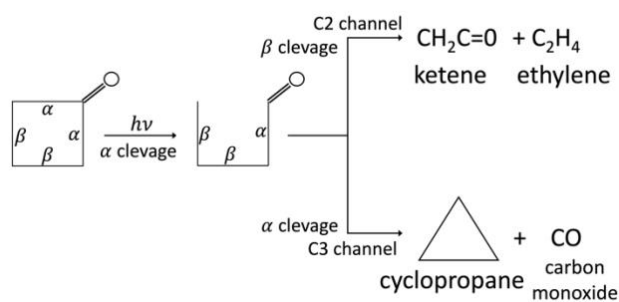

Figure 1:    The schematics of two main photo-pathways of cyclobutanone (CB). Upon photoexcitation, CB forms a diradical intermediate through a C-C bond breaking in the $\alpha$ position. Subsequent cleavage of a $\beta$ bond or the second $\alpha$ bond yields the C2 and C3 photoproducts, respectively.

Diau, Kötting, and Zewail[17] employed femtosecond time-resolved mass spectrometry to explore the CB photoinduced dynamics. They reported that one-photon excitation at 307 nm to the $S_1$ state induces two different dynamical processes associated with the $\alpha$−cleavage in the $S_1$ state. The fast decay component with a time constant of less than 50 fs is mainly responsible for the ultrafast relaxation of the initial wavepacket from the Franck-Condon to the $S_1$ minimum region. On the other hand, the wavepacket moves along the bond-breaking coordinates, generating the photoproducts in the slower 5 ps decay process. They further confirmed using DFT calculations that the CO stretching, CO out-of-plane wagging (pyramidalization), and the ring-puckering modes are the primary nuclear vibrations in the $S_1$ surface, which should transfer the excess energy to the ring-opening modes through IVR to generate the photo-products.

A static theoretical study[21] using CASSCF and MS-CASPT2 methods revealed that the ring-opening mechanism has a barrier of 0.29 eV in the $S_1$ state, which can be easily surmounted by the excess energy of the $S_1$ state after 4.1 eV photoexcitation. After that, the internal conversion to $S_0$ is downhill through an $S_1/S_0$ conical intersection of 3.5 eV energy. This offers a fast pathway that should dominate over an intersystem crossing followed by the ring opening in the $T_1$ state. Ab initio multiple spawning dynamics using SA(2)-CASSCF(12,11)/6-31G* was employed following photoexcitation to the $S_1$ state of CB. The $S_1$ lifetime was estimated to be 484 fs.[22] An average time of 177 fs was estimated for the $\alpha$−cleavage. After reaching the ground state, two comparable bond-breaking mechanisms, involving either a second $\alpha$− or a $\beta$−cleavage, led to the C3 and C2 product formation with a 1.03 C3:C2 ratio.

Using a combination of time-resolved mass spectrometry (TR-MS) and time-resolved photoelectron spectroscopy (TR-PES), the decay of the $n \rightarrow 3s$ Rydberg state ($S_2$) to the $n\pi^\star$ state was characterized[23] by two time constants of 0.08 ps and 0.74 ps. A low-frequency ring-puckering mode of 35 cm$^{-1}$ was suspected to promote the transition from the $S_2$ to the $S_1$ state. A five-dimensional (ring-puckering, CO out-of-plane, CO stretching vibrations, and symmetric and asymmetric $\alpha$ C-C stretchings) vibronic coupling Hamiltonian model was constructed[24] using EOM-CCSD to model cyclobutanone's four lowest singlet electronic states. MCTDH quantum dynamics simulations using this model Hamiltonian predicted a biexponential $S_2/S_1$ internal conversion with time constants of 0.95 ps and 6.32 ps.

Regardless of previous investigations, a complete description following the $S_2$ photoexcitation of cyclobutanone to the end-product formation is yet to be seen. To deal with all this uncertainty, we simulated the photoinduced dynamics using surface hopping methodology combined with different electronic structure methods. These results allow us to discuss the $S_2$ lifetime, the internal conversion mechanisms, the photoproducts, and the dependence of the outputs on the method choice. Gas phase ultrafast electron diffraction (GUED) patterns are provided as a side product to compare the upcoming experimental results.

## 2  Computational Methods

### 2.1  DC-FSSH Dynamics

DC-FSSH is a mixed quantum-classical nonadiabatic dynamics methodology, where a swarm of classical and independent trajectories approximates the nuclear wavepacket evolution through multiple electronic states. Each trajectory is propagated on a single potential energy surface (PES) calculated by solving the time-independent Schrödinger equations for the electrons within the Born-Oppenheimer approximation. During the propagation, a stochastic algorithm,[25] considering the instantaneous probability of nonadiabatic transitions between electronic states, dictates whether the trajectory will hop to another electronic state or stay on the current one. The DC-FSSH algorithm generally needs the electronic energies, force (negative gradient of potential energy), and nonadiabatic couplings from the electronic structure calculations. It is structured so that global energy surfaces are never required, and an on-the-fly approach, computing electronic properties as needed, can propagate the trajectories.

The hop probabilities are computed with a local approximation of the time-dependent Schrödinger equation for the electron and depend on the time-derivative couplings $\sigma_{JL} = \left\langle J \left| \frac{d}{dt} L \right. \right\rangle$, between adiabatic electronic states $J$ and $L$. These couplings can be obtained from the first-order nonadiabatic coupling vector (NACV), $\boldsymbol{h}_{JL} = \left\langle J \left| \frac{\partial}{\partial \boldsymbol{R}} L \right. \right\rangle$, or other approaches such as the Hammes-Schiffer-Tully[26] and local diabatization methods,[27] both relying on wavefunction overlaps.[28] Alternatively, the coupling can be estimated from the PES topography through the time-dependent Baeck-An (TDBA) approach,[29] as demonstrated in Refs.[30–32]

It is well established that the original FSSH algorithm suffers from over-coherence, as state superpositions, eventually born during the time-dependent Schrödinger equation propagation, are never eliminated, yielding an inconsistency between trajectories lying on single electronic states and entangled electronic populations. Different approaches have been proposed to deal with this problem.[33,34] In this work, we apply the simplified decay of mixing (SDM), which exponentially scales down the electronic population of unoccupied states with a decoherence lifetime estimated from the energy gaps and nuclear kinetic energies.[35]

We carried out different sets of DC-FSSH nonadiabatic dynamics using NACVs as obtained by solving the time-independent Schrödinger equation and with approximated TDBA couplings to accelerate the dynamics. The TDBA model used analytical second derivatives from quadratic regression ($\Delta T = 0.4$ fs) and $\delta\eta = 0.1$ a.u. In both cases, the Newtonian equations of motion were integrated using the Velocity Verlet algorithm with a timestep of 0.5 fs. The time-dependent Schrödinger equation was integrated using interpolated electronic quantities between classical steps with a time step of 0.025 fs. The decoherence parameter in SDM was $\alpha = 0.1$ a.u.[35] After a successful hop, the total energy was balanced by rescaling the nuclear velocity in the direction of the NACV when available. In DC-FSSH with TDBA couplings, the nuclear velocity was adjusted in the momentum direction. In this case, to decide whether hopping from a lower into a higher electronic state was allowed or frustrated, we used the reduced kinetic energy reservoir discussed in Ref.[31] to ensure size-extensivity. In the case of a frustrated hopping, the momentum direction was not changed.

The initial geometries for the dynamics were randomly sampled from a marginal Wigner distribution for the quantum harmonic oscillator in the electronic ground state equilibrium geometry. Initial momentum was selected to have the total energy of each trajectory equating to the zero point energy.[36] The initial energy excitation and number of trajectories in each set are shown later in Section 3.2.

We have sampled one thousand initial conditions for each one of the electronic structure methods. These geometries were used to simulate the absorption spectra of CB in the gas phase with the nuclear ensemble approach.[37]

We performed several sets of DC-FSSH nonadiabatic dynamics simulations in the gas phase, considering cyclobutanone's lowest three singlet electronic states ($S_0$, $S_1$, and $S_2$). For each of these sets, we used different electronic structure methods and nonadiabatic coupling approaches, as follows:

- Set-1: DC-FSSH with NACV and multiconfigurational self-consistent field (MCSCF) [MCSCF-NACV];

- Set-2: DC-FSSH with TDBA and MCSCF [MCSCF-TDBA];

- Set-3: DC-FSSH with TDBA and linear-response time-dependent density functional theory (TDDFT) for excited states and DFT for the ground state [TDDFT-TDBA];

- Set-4: DC-FSSH with NACV and multireference configuration interaction (MRCI) based on the semiempirical orthogonalization and dispersion corrections method ODM3 [MNDO-NACV].

- We also report our experience attempting to use DC-FSSH with TDBA in combination with either coupled cluster with approximated second-order (CC2) or algebraic diagrammatic construction to the second-order [ADC(2)].

The trajectories were initiated on the $S_2$ electronic state in all the simulations. Since the molecule has enough excess energy to surmount the fast ring-opening and deactivation channels in the $S_1$ state, we can safely neglect[21] any possible role of triplet states having intersystem crossing on nanosecond timescales[17] in this photoinduced dynamics.

Dynamics based on active-space methods, such as CASPT2, MRCI, MCSCF, or CASSCF, are usually unstable due to orbital exchange between subspaces when the molecule explores the configurational space.[38] This instability commonly implies total energy conservation failure. Although our MCSCF reduces this problem by providing enough flexibility to describe even dissociated structures, we still observed active space instabilities in Set-1 and Set-2 trajectories with MCSCF near $S_2/S_1$ conical intersection featuring α-cleavage. As a result, trajectories entering the crossing region via α cleavage underwent a sudden change of potential energy, leading to total energy change. To overcome this problem, we leveraged these trajectories by imposing moderately loose energy conservation thresholds in our dynamics. This loosening protocol can be reasoned by safely assuming that the trajectories passed the CI region at the crashing point and effectively funneled down to the $S_1$ state. Consequently, the trajectory continued steady propagation with the altered total energy.

We also followed the same loose energy conservation protocol in Set-4 simulations with MRCI/ODM3, which also suffers from similar active-space instabilities. On the other hand, because TDDFT is not subjected to such a problem, Set-3 trajectories were propagated with strict energy conservation criteria.

Due to the limitation of TDDFT in dealing with the multireference ground states, trajectories were stopped whenever the energy difference between $S_1$ and $S_0$ states dropped below 0.1 eV. In several cases, however, the trajectories stopped with a small $S_1/S_0$ energy gap due to convergence errors in DFT calculations. In these cases, the trajectories were considered as they reached the energy difference criterion. The time at which this threshold was attained was considered as the $S_1/S_0$ internal conversion time.[39,40] A few trajectories succeeded in hopping to the ground state and were continued for a few fs in this state. All trajectories were restarted in the $S_0$ at the DFT level using the velocities and positions from the last point they reached with the TDDFT dynamics.

The details of the electronic structure methods are given in the following sections. The initial conditions, absorption spectra, and dynamics simulations were done using the Newton-X CS suite of programs (version 2.7).[36]

## 2.2   Electronic Structure Methods

Set-1 and Set-2 dynamics simulations were carried out with the state-averaged MCSCF method based on the graphical unitary group approach (GUGA) using Columbus software (version 7.2.2, Sept. 2022)[41,42] utilizing analytic energy gradients and nonadiabatic coupling procedures as described in Refs.[43–45] The MCSCF wave function was constructed within an extended general valence bond (GVB) perfect pairing (PP) multiconfigurational approach,[46–48] combining a complete active space (CAS) with PP subspaces. Such an MCSCF scheme yields a compact representation of the wavefunction light enough to allow extended dynamics simulations but still robust enough to describe the excitation and occasional dissociation. This type of wavefunction has been successfully used previously for the photodynamics of ethylene,[49] azomethane,[45,50] and in extended form for the accurate calculation of molecular bond lengths.[51]

The active space in the MCSCF wavefunction comprises one complete active space (CAS) and five PP subspaces. The CAS subspace contains four electrons in four orbitals [CAS(4,4)], including a nonbonding lone pair orbital $n$, a 3s Rydberg orbital originating at the oxygen atom, and the C=O $\pi$ and $\pi^*$ orbitals. (The molecular orbitals are presented in Section S1 of the Supplementary Material (SM). This part of the wavefunction was designed to describe the electronic ground state and the two excited states ($n\pi^\star$ and $n \rightarrow 3s$). Each of the five PP subspaces contains two electrons in two orbitals, allowing only double occupation in each orbital. These orbital pairs are the four C-C σ and σ* orbitals and the C-O σ and σ* orbitals. The four PP C-C subspaces allow independent bond breaking, enabling the description of the opening of the CB ring. The remaining electrons were kept in doubly occupied orbitals. This MCSCF wavefunction provides a smooth transition to the ground state and a continuation with the hot ground state dynamics. Besides that, it has only 640 configuration state functions (CSFs). A schematic representation of the MCSCF wavefunction is

MCSCF(14,14) = 4[PP$_{CC}$] × [PP$_{CO}$] × [CAS(4,4)$_{n,\pi,\pi^*,3s}$]

Three electronic states were computed with a state averaging procedure (SA-3). It was used with the cc-pVDZ basis set for hydrogen, the aug-cc-pVDZ for carbon, and the d-aug-cc-pVDZ for oxygen.[52] This hybrid basis set describes all orbitals well and delivers good excitation energies at reasonable computational demand. Such a balance is essential given the vast number of MCSCF calculations in the on-the-fly dynamics propagation.

Minima on the crossing seam were calculated with MCSCF using COLUMBUS's analytic energy gradient and nonadiabatic coupling vector features. These optimizations were performed with a modified version of GDIIS as described using natural coordinates.[44,53,54] For simplicity, these minima on the crossing seam are denoted conical intersections (CIs) in the following text.

The Set-3 simulations were performed with TDDFT followed by DFT dynamics in the ground state using the CAM-B3LYP functional.[55] The basis set consists of cc-pVDZ with an augmented set of $1s1p1d$ diffuse functions for the oxygen atom. All the DFT (ground state optimization, frequency calculations) and TDDFT calculations were done using ORCA v5.0.4.[56] Due to the limitation of the TDDFT method in dealing with the multireference ground states, trajectories were stopped whenever the energy difference between S$_1$ and S$_0$ states dropped below 0.1 eV. In several cases, however, the trajectories stopped with a small S$_1$/S$_0$ energy gap due to convergence errors in DFT calculations. In these cases, the trajectories were considered as they reached the energy difference criterion. The time at which this threshold was attained was considered as the S$_1$/S$_0$ internal conversion time.[39,40] A few trajectories succeeded in hopping to the ground state and were continued for a few fs in this state. All trajectories were restarted in the S$_0$ at the DFT level using the velocities and positions from the last point they reached with the TDDFT dynamics.

In Set-4, we employed the multireference configuration interaction with single and double excitations (MRCI) based on ODM3 as implemented in the MNDO program.[57,58] The MRCI calculation employed an active space of 3 occupied (6 electrons) and three unoccupied orbitals. The number of references was automatically selected so that these configurations contributed at least 85% of the weight in all three requested CI roots. Energy gradients and NACV were computed at the same level.

Additional calculations were performed with the resolution-of-identity (RI) CC2 and ADC(2) using the Turbomole v7.6 program.[59] For these calculations, the cc-pVDZ basis set was employed for hydrogen and carbon, and the d-aug-cc-pVDZ was used for the oxygen.

## 3    Results and Discussion

### 3.1    Potential Energy Topography

The ground state (S$_0$) equilibrium geometry of the gas phase CB computed by all methods is nearly planar. At the S$_0$ minimum, the first excited singlet state (S$_1$) has a dark $n\pi^\star$ character, and the second excited singlet state (S$_2$) is a bright Rydberg state originating from a $n \rightarrow 3s$ transition. MCSCF shows an S$_1$ minimum nonplanar with the ring puckering dihedral angle of 14°, and the out-of-plane carbonyl dihedral angle is 146° (Figure 2). The carbonyl bond length increases from 1.2 Å in the ground state to 1.4 Å in the S$_1$ $n\pi^\star$ state. Also using MCSCF, the S$_2$ minimum is again planar and similar to the ground state minimum. The S$_1$ and S$_2$ minimum energy structures retain their electronic character initially assigned at the S$_0$ equilibrium geometry. Figure 2 shows the optimized minimum energy structures and conical intersections computed with MCSCF. The $\alpha$ and $\beta$-carbonyl C-C bond lengths, as well as the ring-puckering dihedral (C-C-C-C) and the carbonyl out-of-plane dihedral (O-C-C-C) angles, are shown. These quantities computed with DFT/TDDFT and MRCI/ODM3 are shown in the SM Section S2. The geometries calculated with different electronic structure methods compare well with each other and with the previously reported ones (see also SM S2).

Table 1 collects the excitation energies and oscillator strengths of the first and second singlet excited states at the ground state minimum by different methods and compares them with previously computed values. Taking EOM-CCSD/cc-pVTZ[24] as the reference result, our MCSCF underestimates the vertical excitation into S$_1$ by about 0.2 eV and overestimates the S$_2$ energy by approximately the same amount. This can be rationalized by the unbalanced account of dynamic electron correlation in MCSCF.

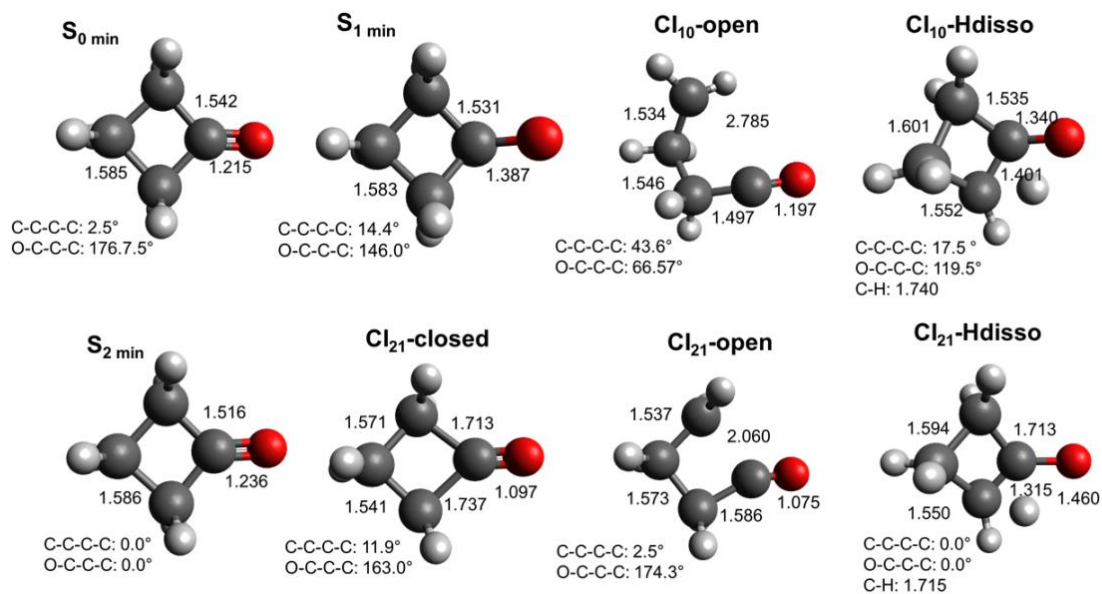

Figure 2: Geometries of minima and minima on the crossing seam (conical intersections) computed with MCSCF.

Table 1: Excitation energies (eV) and oscillator strengths (given in parentheses) at the $S_0$ minimum with MCSCF, TDDFT, and MRCI/ODM3. Previously reported computational results are shown as well.

| | Methods | $S_1$ ($n\pi^*$) | $S_2$ ($n \to 3s$) |
|---|---|---|---|
| | MCSCF | 4.24 (0.000) | 6.77 (0.012) |
| Present | TDDFT | 4.28 (0.000) | 6.44 (0.043) |
| | MRCI/ODM3 | 4.09 (0.000) | 6.35 (0.306) |
| | SA(2)-CAS(12,11)/6-31G* [Ref.22] | 4.39 | |
| Previous | MS-CASPT2//CASSCF(8,7)/6-31+G* [Ref.21] | 4.10 | |
| | EOM-CCSD/cc-pVTZ+1s1p1d [Ref.24] | 4.45 | 6.60 |

The excitation energies calculated at different critical points along the reaction coordinate, including $S_2$ and $S_1$ minima and several conical intersections, are given in Table 2. At the $S_1$ minimum, the MCSCF $S_1$ excitation energy matches perfectly with the EOM-CCSD result. Since the $S_2$ minimum structure is very similar to the ground state minimum, the electronic energies are also near their corresponding Franck-Condon values.

Conical intersections were computed only using MCSCF. We identified three $S_2/S_1$ conical intersections (CI) types with distinctly different structural and electronic features. (see Table 2). An $S_2/S_1$

CI with an $\alpha$ ring-opening and planar structure (CI$_{21}$-open) lies 0.1 eV above the S$_2$ minimum. The diabatic character of the crossing electronic states is S$_2(\sigma_{CC} \to 3s)$ and S$_1(\sigma_{CC} \to \pi^*)$. Another S$_2$/S$_1$ CI with a closed and slightly puckered ring (CI$_{21}$-closed) exists 0.5 eV above the S$_2$ minimum. In this CI, the diabatic character of the crossing electronic states is the same as at the S$_2$ minimum, S$_2(n \to 3s)$/S$_1(n \to \pi^*)$. A third S$_2$/S$_1$ CI, featuring a dissociated hydrogen initially bonded to an $\alpha$ carbon atom, lies 0.6 eV above the S$_2$ minimum (CI$_{21}$-Hdisso). The electronic transitions associated with this CI are S$_2(n \to \sigma_{CH}^\star)$/S$_1(\pi_{CO} \to \sigma_{CH}^\star)$. The structures, electronic states, and molecular orbitals of the CIs can be visualized in SM S3.

Two S$_1$/S$_0$ CIs were found: one showing a $\alpha$-cleavage (CI$_{10}$-open) and the other with a dissociated-H (CI$_{10}$-Hdisso). CI$_{10}$-open lies 3.98 eV above the ground state minimum (see Table 2), 0.46 eV higher than the previously reported MS-CASPT2 estimate.[21] It corresponds to an S$_1(\sigma_{CC} \to \pi_{CO}^\star)$/S$_0$(closed shell) crossing. CI$_{10}$-Hdisso lies at 6.1 eV, with an S$_1(n \to \sigma_{CH}^\star)$/S$_0$(closed shell) crossing. Both CIs are energetically accessible after the initial photoexcitation into S$_2$.

Table 2: Vertical excitation energies (eV) at the S$_1$ and S$_2$ minima and optimized conical intersections calculated with MCSCF. The energies are compared with the previously reported values with different levels of electronic structure theories.

| | S$_1$ min | S$_2$ min | CI$_{21}$-open | CI$_{21}$-closed | CI$_{21}$-Hdisso | CI$_{10}$-open | CI$_{10}$-Hdisso |
|---|---|---|---|---|---|---|---|
| S$_0$ | 1.22 | 0.07 | 3.77 | 2.39 | 6.80 | 3.98 | 6.118 |
| S$_1$ | 3.20<br>3.84[#]<br>3.50[$]<br>3.22[##] | 4.13 | 6.783 | 7.196 | 7.31 | 3.98<br>2.90[#]<br>3.52[$] | 6.123 |
| S$_2$ | 8.14 | 6.70 | 6.785 | 7.197 | 7.31 | 7.29 | 11.19 |

[#]SA(2)-CAS(12,11)/6-31G*;[22] [##]CASSCF(10,8)/6-31G(d);[17] [$]MS-CASPT2//CASSCF(8,7)/6-31+G*;[21] [$$]EOM-CCSD/cc-pVTZ+1s1p1d.[24]

Considering the topography of the PESs computed by MCSCF (see Figure 3), we may hypothesize that after photoexcitation into the S$_2$ state (6.77 eV), CB will relax to the S$_2$ minimum (6.70 eV). From there, it may preferentially convert to S$_1$ through the CI$_{21}$-open, which is energetically nearby (Pathway 1). Then, CB would relax on S$_1$ via the ring-opening mechanism. Nevertheless, we cannot discard competing mechanisms where CB would convert to S$_1$ at the other two CI types by surmounting barriers of about 0.5 eV energy, which is well within the available energy after photoexcitation (Pathways 2 and 3). Since the S$_1$ minimum is a puckered closed-ring structure, the molecules relaxing from the S$_2$ state via the $\alpha$-





cleavage and H-dissociated CIs would not likely relax to the $S_1$ minimum. Instead, they may undergo rapid internal conversions to the ground state via $S_1/S_0$ CIs.

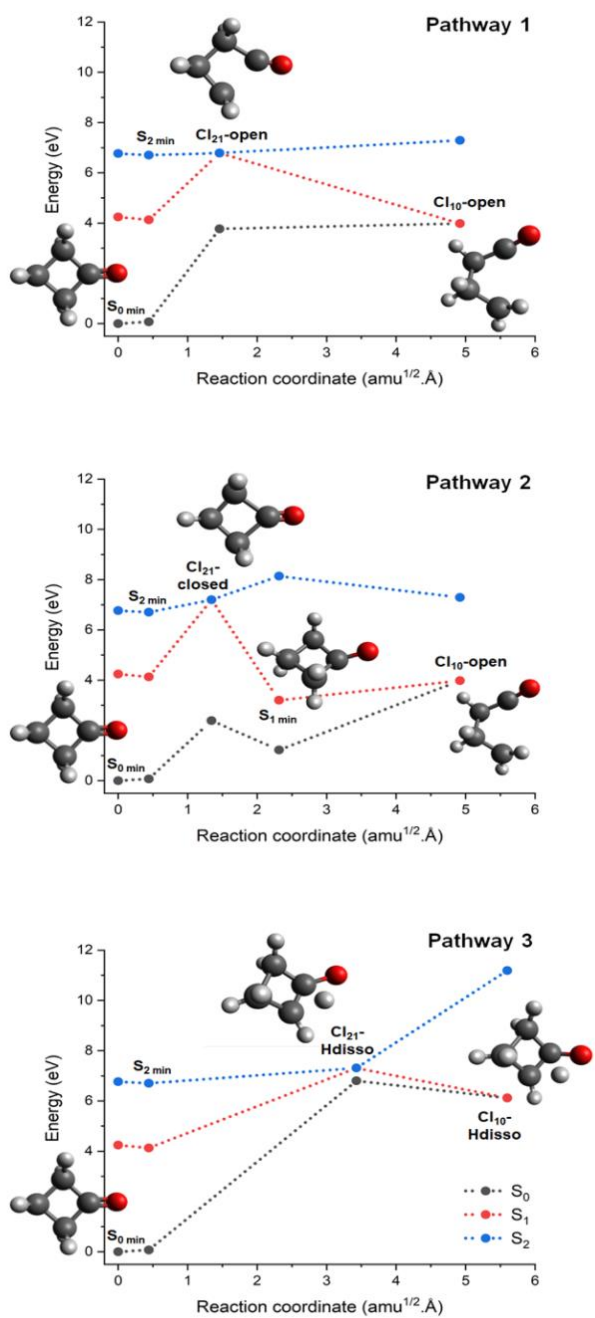

Figure 3:     Topography of the $S_0$, $S_1$, and $S_2$ PESs involved in the dynamics calculated with MCSCF.





## 3.2    Absorption Spectra

Figure 4 shows the CB absorption spectrum in the gas phase up to the $S_2$ ($n \rightarrow 3s$) band computed with the nuclear ensemble approach. Compared to the experiments, MCSCF predicts an $S_2$ ($n \rightarrow 3s$) band blue-shifted by 0.32 eV and underestimates the oscillator strength by a factor of 4.[60] MRCI/ODM3 strongly overestimates the oscillator strengths, likely due to the compactness of its wavefunction representation, but delivers reasonable $S_2$ band energies. TDDFT/CAM-B3LYP delivers an excellent $S_2$ ($n \rightarrow 3s$) band prediction thanks to the diffuse basis set and the range-separated functional. The $S_1$ band intensities of the experimental,[60] MCSCF, and TDDFT spectra are too small to appear on this scale. They are shown in the SM S4.

The proposed upcoming experiment will be conducted by exciting the $S_2$ state with a 200 nm (6.2 eV) laser. Comparing the shape and the position of the simulated $S_2$ band with the experimental one, we randomly sampled initial conditions for our dynamics simulations constrained to the $6.65 \pm 0.03$ eV excitation energy window (shaded area marked by A in Figure 4) for MCSCF-NACV (Set-1) and MCSCF-TDBA (Set-2) simulations. On the other hand, an energy window of $6.20 \pm 0.05$ eV (shaded area marked by B) is used to sample initial conditions for TDDFT-TDBA (Set-3) and MRCI/ODM3-NACV (Set-4) simulations. Hence, we have 93 independent trajectories for Set-1, 88 for Set-2, 78 for Set-3, and 190 for Set-4.

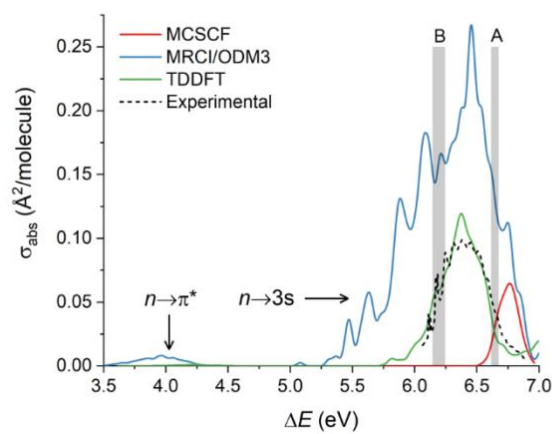

Figure 4:    Simulated and experimental[60] absorption spectra of cyclobutanone in the gas phase, emphasizing the $S_2$ ($n \rightarrow 3s$) band. The shaded areas A and B indicate energy windows from where initial conditions for dynamics were selected (see text for details).





### 3.3 Population Time Evolution

Figure 5 shows the average trajectory occupation (fraction of trajectories in a particular state in a given timestep) in the three states as a function of time for different sets of simulations. We fitted the $S_2$ occupation with a mono-exponential decay function, $f(t) = e^{-(t-t_{lag})/t_{exp}}$, where $t_{exp}$ is the exponential decay time constant of the $S_2$ state and $t_{lag}$ refers to the time after the $S_2$ occupation starts to decay. The lifetimes is $\tau = t_{lag} + t_{exp}$. These parameters are tabulated in Table 3.

Table 3: $S_2$ lifetimes ($\tau$), lag times ($t_{lag}$), and exponential decay times ($t_{exp}$) as obtained from different sets of simulations by a mono-exponential decay fit.

| Simulations | $t_{lag}$ (ps) | $t_{exp}$ (ps) | $\tau$ (ps) |
|---|---|---|---|
| MCSCF-NACV (Set-1) | 0.3 | 8.7 | 9.0 |
| MCSCF-TDBA (Set-2) | 0.3 | 10.1 | 10.4 |
| TDDFT-TDBA (Set-3) | 0.00 | 0.40 | 0.4 |
| MRCI/ODM3-NACV (Set-4) | 0.005 | 0.035 | 0.04 |
| ADC(2)-TDBA | 0.0 | 0.9 | 0.9 |

With MCSCF (Set-1 and Set-2), the $S_2$ state starts to decay after 300 fs, and its occupation reduces to ~30% after 10 ps. The predicted lifetimes of the $S_2$ state for Set-1 and Set-2 simulations are 9 and 10 ps, respectively. It is also interesting to note that the $S_1$ state is practically never populated. Once the trajectories hop down from the $S_2$ to the $S_1$ state, they immediately undergo another hopping event to the ground state. The mean time difference between the two successive hops (i.e., the difference between $S_2/S_1$ and $S_1/S_0$ hopping times) is 16 fs. Hence, the trajectories never relax to the $S_1$ minimum, as predicted by considering the topography of the PESs. The occupation profiles obtained by MCSCF-NACV (Set-1) and MCSCF-TDBA (Set-2) match closely considered the statistical uncertainty. The bootstrapped 95% confidence interval (with $10^5$ resamplings) is 1 and 2 ps for Set-1 and Set-2, respectively. Therefore, these results validate using approximated TDBA couplings.

TDDFT-TDBA (Set-3) simulations show a starkly different result than the MCSCF dynamics. It predicts a much slower population transfer compared to Set-4. After photoexcitation to the $S_2$ state, CB immediately starts to decay (having no lag time) to the $S_1$ state. The population is entirely transferred to





the ground state, yielding an $S_2$ lifetime of 0.4 ps. The population starts accumulating in the $S_1$ state from the beginning and attains a maximum of 35% around 200 fs. Afterward, it exponentially decays to the ground state in the $S_1$ state population.

MRCI/ODM3-NACV dynamics have features similar to TDDFT-TDBA but a ten times faster timescale. In this case, the $S_2$ population completely decays to the ground state within 200 fs, predicting its lifetime as 40 fs. The growth of the $S_1$ population reaches a maximum value of 60% around 50 fs and then exponentially decays to the ground state.

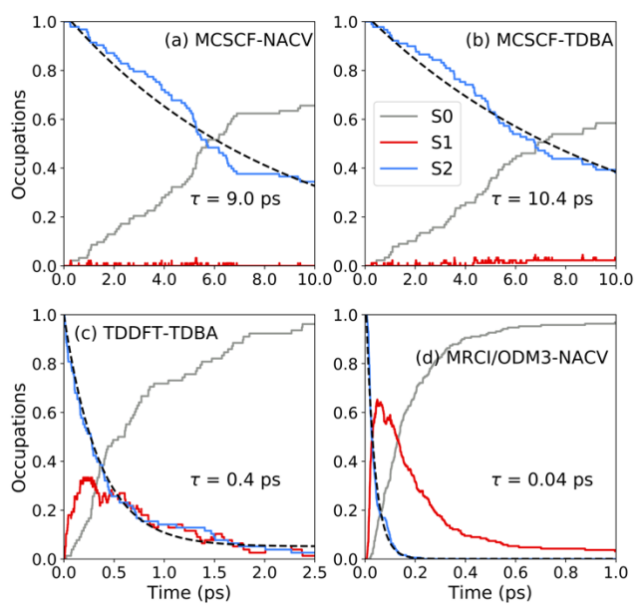

Figure 5: The time evolution of average occupations of the three states for different sets of simulations.

Our attempts to run the dynamics using CC2 and ADC(2) methods did not work well, as expected. Using CC2, most of the trajectories (77 out of 91) were finished too early due to convergence problems in the electronic structure calculations caused when $S_2$ and $S_1$ were energetically close. On the other hand, the trajectories using ADC(2) were less problematic in terms of convergence: only 8 out of the 80 trajectories stopped due to eigenvalue convergence failures, already in the $S_1$ state. However, although the hopping geometries from $S_2$ to $S_1$ were reasonably described, the $S_1/S_0$ hoppings were not correctly predicted, as most of the geometries show an abnormal stretching of the C=O bond above 1.5 Å (see SM S5). This problem with ADC(2) has been recently discussed in Ref.[10] The fitting of the $S_2$ fraction of population decay provided this state a lifetime of 0.9 ps.





### 3.4 Internal Conversion Mechanisms

As speculated from the PES topography (see Section 3.1), indeed, after photoexcitation to the $S_2$ state, CB may convert to $S_1$ via one of the three different pathways related to the three $S_2/S_1$ conical intersections (CI$_{21}$-open, CI$_{21}$-closed, and CI$_{21}$-Hdisso). Figure 6 depicts the distribution of the $S_2$-$S_1$ energy gaps at the $S_2/S_1$ hopping geometries near each of these conical intersections using different methods.

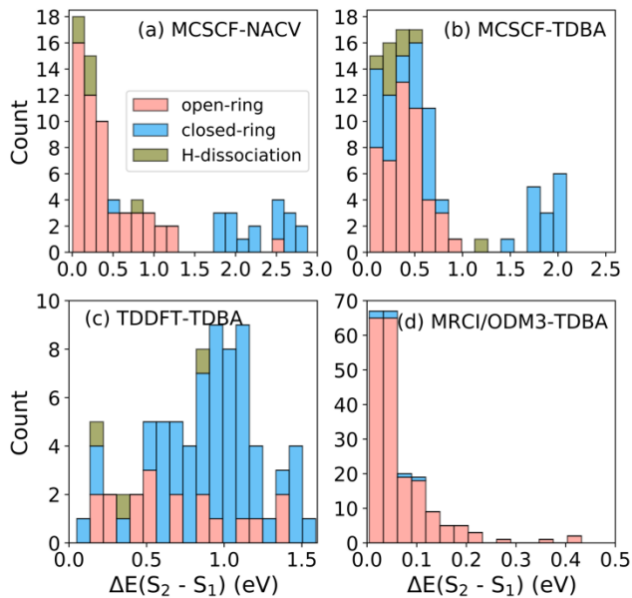

Figure 6: Distribution of energy gaps between $S_2$ and $S_1$ states at the $S_2/S_1$ hopping geometries for the four sets of simulated trajectories.

In MCSCF-based dynamics (Set-1 and Set-2), most of those trajectories (47%) approached the CI$_{21}$-open via the ring-open mechanism. The energy gaps of hopping geometries peaking at lower energy signifies attaining the CI$_{21}$-open region in the MCSCF-NACV dynamics [see panel (a) of Figure 6]. A few trajectories (5%) went uphill, surmounted the higher-energy CI$_{21}$-Hdisso, and converted to the $S_1$ state. In MCSCF-TDBA dynamics, we also observe similar $S_2/S_1$ internal conversion mechanisms. The distribution of the ring-puckering angles at the hopping geometries is given in SM S6. The $S_2/S_1$ hoppings mainly occur with small ring-puckering geometries. However, some hoppings occur around a 60° C-C-C-C dihedral angle, revealing that the $S_2/S_1$ crossing seam extends into twisted configurations. A final observation about MCSCF simulations: 30% of the trajectories remained unreactive in the $S_2$ state until





the end of 10 ps simulation time. This can also be seen in the remaining $S_2$ population in panels (a) and (b) of Figure 5.

In Set-3 of TDDFT-TDBA dynamics, most of the trajectories (73%) underwent closed-ring $S_2/S_1$ hops with minor and moderate energy gaps. In this case, the ring-opening mechanism (23%) at the $S_2$ state seems not dominant.

In Set-4 of MRCI/ODM3-TDBA dynamics, almost all trajectories (92%) were converted from $S_2$ to the $S_1$ state via the ring-opening pathway.

At the $S_1$ state, most of the trajectories (57% of the total and 86% of the $S_1$ state) were further converted to the ground state with ring-opened structures in the MCSCF-NACV (Set-1) simulation (see Figure 7). Only 7% of the total trajectories (11% of trajectories in the $S_1$ state) reached the ground state via the $CI_{10}$-Hdisso channel. It means that the closed-ring trajectories in the $S_1$ state also underwent a C-C bond cleavage to reach the $S_0$ state. The other simulations with MCSCF-TDBA, TDDFT-TDBA, and MRCI/ODM3-NACV also had the ring-open as the dominant path for the $S_1/S_0$ internal conversion.

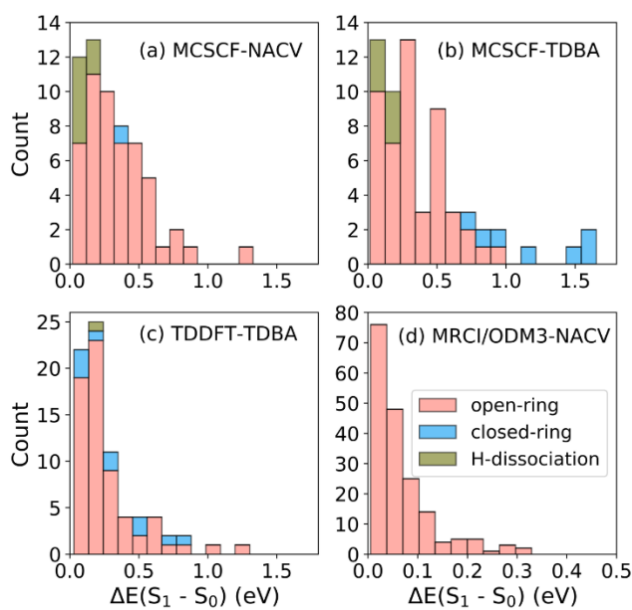

Figure 7:     Distribution of energy gaps between $S_1$ and $S_0$ states at the $S_1/S_0$ hopping geometries for the four sets of simulated trajectories.

A final observation should not be missed here: 16% of MCSCF-NACV trajectories attaining closed-ring structures show $S_2/S_1$ hoppings with energy gaps between 2 and 3 eV (See the top panels in Figure 6). At first, it seems unusual but not impossible with the instantaneous probabilities in the





framework of FSSH dynamics in the weak coupling regime.[61] The trajectories spend a long enough time with moderate energy gaps, with low transition probability (small NACV) but with many hopping opportunities until a small enough random number triggers the stochastic hopping. Similar high-energy hoppings are also observed in Set-2 simulation with MCSCF-TDBA. However, the $S_2/S_1$ hoppings tend to occur at smaller energies with TDBA than with NACV. This is an artifact of the way TDBA was set. These approximated couplings were computed only when the energy gaps were smaller than 1 eV. Thus, the hops were constrained to be below this threshold. However, increasing the threshold to 2 eV, the $S_2/S_1$ CI-closed hoppings occur at higher energies, as with NACV. We show these high-energy hoppings for MCSCF-TDBA in Figure 6, too. These large-gap hoppings in the $S_2/S_1$ closed structures were also observed with TDDFT (Figure 6) and ADC(2) (SM S5).

### 3.5    Photoproducts

Working with different data sets and trajectories finishing at different times, it was crucial to fix a protocol to classify the photoproducts in terms of geometrical parameters and the time to observe their geometries. Based on the trajectory survival time in $S_0$, we chose to classify the photoproducts 200 fs after the $S_1/S_0$ hopping. In a few cases, we took the last geometry, when the trajectories did not survive till that pre-defined time due to the failure of electronic structure convergence, usually due to dissociation events.

To check for single-bond cleavage, we defined parameters based on $r_{mean} + 3\sigma$ values for each bonded pair by fitting the particular bond distances with a Gaussian distribution, considering all the molecular geometries obtained in the dynamics. In this fitting, $r_{mean}$ is the mean and $\sigma$ is the standard deviation. If the atomic distances exceeded $r_{mean} + 3\sigma$, we considered it a single bond cleavage.

After that, the C2 products, which involve double bond cleavage, were assigned if a $\beta$ and an $\alpha$ CC distance were greater than 2.5 Å. Similarly, the C3 products were assigned if both $\alpha$ CC distances were greater than 2.5 Å. Thus, C3 products contain CO + cyclopropane or propene molecules. Although we observed propene conversion in cyclopropane in a few trajectories propagated already in the ground state, we did not classify them as different channels. Analogously, C2 products containing ethylene + ketene molecules or ethylene + CO + $CH_2$ fragments were counted in the same channel. We also verified at times longer than 200 fs that the C3 and C2 products remained dissociated.

We observed an interesting mechanism in very few trajectories. A hydrogen atom was detached from an $\alpha$- carbon position and followed a roaming pathway toward the oxygen atom. Since this happened





only in a few trajectories, we are not in a position to predict the roaming mechanism confidently. Thus, they were generally classified as H-dissociation.

Following this protocol, the photoproducts were classified as a closed ring, open ring (single bond cleavage), C2 ($\alpha$ and $\beta$ cleavages), C3 (two $\alpha$ cleavages), and H dissociation. The amounts of each obtained in the different sets of simulations are tabulated in Table 4.

Table 4:    Amount (%) of different photoproducts obtained after the trajectory spends 200 fs in the ground state.

| Simulations | closed-ring | open-ring | C3 products | C2 products | H-dissociation |
|---|---|---|---|---|---|
| MCSCF-NACV (Set-1) | 0 | 20 | 32 | 8 | 6 |
| MCSCF-TDBA (Set-2) | 7 | 21 | 28 | 1 | 7 |
| TDDFT-TDBA (Set-3) | 49 | 22 | 5 | 19 | 5 |
| MRCI/ODM3-NACV (Set-4) | 3 | 23 | 28 | 46 | 0 |

Set-1 and Set 2 do not sum to 100% because this table does not include trajectories that remained excited at the end of the simulations.

In the MCSCF-NACV and MCSCF-TDBA dynamics (Set-1 and Set-2), about 32% of the CB molecules remained in the $S_2$ state after 10 ps of simulation, showing no internal conversion to lower states. On the contrary, in the TDDFT-TDBA and MRCI/ODM3-TDBA dynamics (Set-3 and Set-4), the $S_2$ state is entirely converted to $S_0$ within the simulation time.

The presented data encapsulates the outcomes of molecular simulations conducted across various computational methodologies, offering insights into the diverse chemical pathways and products emerging from distinct electronic structure methods. In the following analysis, given the number of trajectories in each set, we consider statistical uncertainty of about ±10% for MCSCF and TDDFT simulations and ±5% for MRCI/ODM3.

In the MCSCF-NACV set, a notable prevalence of open-ring and C3 products is observed. MCSCF-TDBA simulations exhibit a similar trend within the statistical uncertainty, once more confirming the excellent performance of TDBA-approximated couplings. Thus, both MCSCF sets suggest a propensity toward CO elimination in the ground-state photoproducts.





The TDDFT-TDBA simulations portray a distinctive distribution profile characterized by a substantial predominance of closed-ring products alongside notable contributions from open-ring and C2 channels. Therefore, different from MCSCF, TDDFT predicts a predominance of ethylene formation.

The MRCI/ODM3-NACV set exhibits another pattern, characterized by a pronounced abundance of C2 products and a notable representation of C3 species while exhibiting minimal closed-ring occurrences. This distinctive distribution predicts equally large amounts of CO and ethylene eliminations.

## 3.6 Electron Diffraction Spectra

We simulated the GUED signal using the independent atom model (IAM) as prescribed by Centurion et al.[7] This model assumes that the electron arrangement within a molecule mimics that of independent atoms with no interactions among them. As a result, we can safely work with the predetermined atomic force factors (AFFs) to describe the arrangements. However, the approximation overlooks the alterations in electron distribution resulting from the formation of chemical bonds. The working equations to calculate the signals can be seen in SM S7.

The elastic scattering signal is the incoherent sum over an ensemble of all the molecular structures obtained from the nonadiabatic dynamics simulation. To calculate the time-dependent real-space pair-distribution function ($\Delta PDF$), we used a range of 0 to 12 Å$^{-1}$ as the integration limits of the momentum transfer (s) and a damping factor $\alpha = 0.03$ Å$^2$. We used a locally modified version of the "Diffraction_simulation" code,[62] originally developed by Wolf et al., to simulate electron diffraction signals. The AFFs were taken from the git repository (without verification) and were simulated using the ELSEPA program.[63] The GUED diffraction patterns obtained from different sets of simulations are displayed in Figure 8.

Figure 8a shows the PDF signal for the ground state geometry and the most critical contribution of specific atomic interactions to the signal. It is justified to use the same reference to evaluate all sets of simulations, given how well the signal of the initial conditions matches (see section SM S8). In all the methods, it is possible to identify a loss of signal at the distances of 1.5, 2.5, and 3.0 Å associated with the CC and CH distances, suggesting an increase in those distances. It is also possible to see a signal gain around 2 Å, particularly in the MCSCF sets, implying a relaxation of the molecule associated with longer CC bonds.





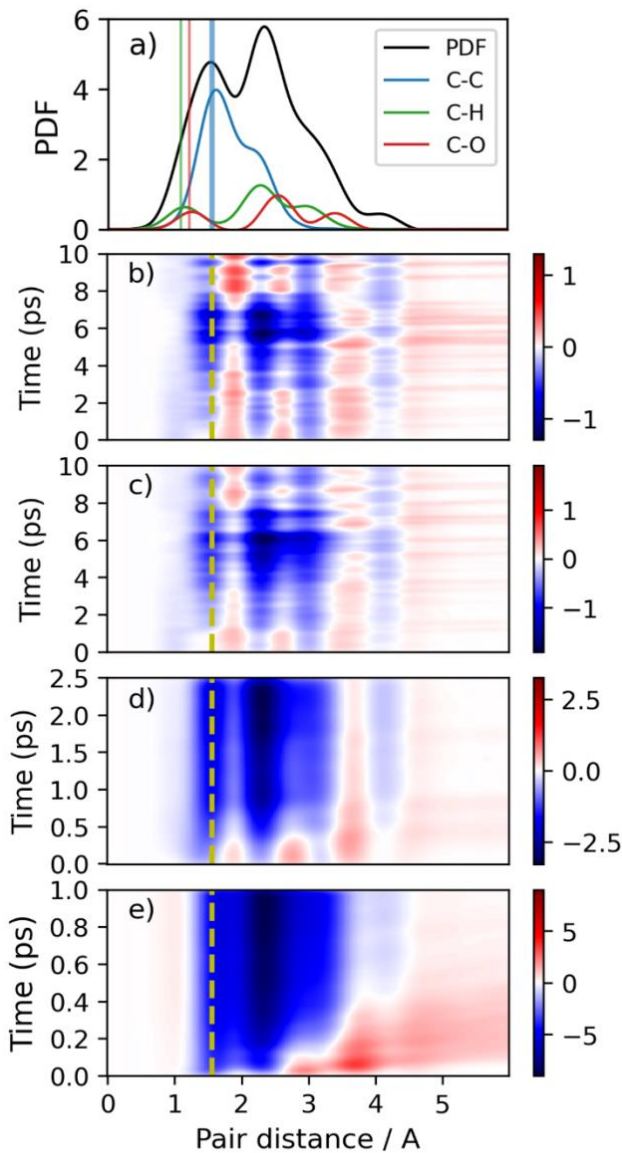

Figure 8: GUED patterns for different sets of simulation, a) shows the PDF for the ground state geometry and the contribution of different bonds. Vertical lines indicate the equilibrium distance; b-e) shows the ΔPDF for the dynamics with different methods, convoluted with 150 fs FWHM. b) MCSCF-NACV, c) MCSCF-TDBA, d) TDDFT-TDBA, e) MRCI/ODM3-NACV. The yellow dashed line marks the $\alpha$ and $\beta$ CC bond distances of the equilibrium geometry of CB.

In the MCSCF sets, this initial profile is somewhat conserved during the whole simulation, caused by a combination of factors. Most importantly, the electronic structure used showed some instability in treating the system after hops, which led to the trajectories stopping after some time, varying from a few





hundred femtoseconds to a few picoseconds. This means that at any given time, most of the trajectories are still in the relaxed structure trapped in the $S_2$, and only a few trajectories yield dissociation. This is complemented by the relatively long timescale of the simulated process, which causes the hops of different trajectories to happen at distant times from each other, leading to the visible horizontal lines at longer pair distances [panels a) and b) of Figure 8].

The calculations with TDDFT show an intermediate timescale associated with more stable trajectories that survived longer after hop. The first difference is in the fast disappearance of the positive signal around 2 Å and smoother gain in signal at longer pair distances.

Finally, MRCI/ODM3 also shows a smooth signal due to the short decay time and the higher number of trajectories allowed by this inexpensive method. The signal also starts with a relaxation phase with signal loss at 1.5, 2.5, and 3.0 Å and an almost unnoticeable signal gain around 2 Å. In this case, however, this relaxation is almost instantaneously followed by the dissociation process, which can be seen by the deep negative signal from 1 to 2.5 Å and a strong positive signal above 3 Å already at 100 fs. In this simulation, almost all dissociation has already occurred in the first half of the dynamics, after which the signal at longer distances appears in lighter red.

## 4    Conclusion

In this work, we explored the dynamics of cyclobutanone after photoexcitation to the $S_2$ ($n \rightarrow 3s$) Rydberg state using DC-FSSH nonadiabatic dynamics combined with different routine electronic structure methods, namely, MCSCF, TDDFT, MRCI/ODM3, CC2, and ADC(2). We also tested DC-FSSH using full and approximated nonadiabatic couplings.

We characterized the $S_2$ lifetime, internal conversion mechanisms, and photoproducts. The results radically depend on the electronic structure method used to propagate the dynamics, as summarized in Table 5

MCSCF predicts a long-lived $S_2$ state, with internal conversions via the ring-open mechanism and predominantly following the C3 channel. Using approximated couplings with TDBA does not change this picture. These outcomes contrast with the TR-PES results from Ref.[23], which reported sub-ps $S_2$ lifetime with the dominance of the C2 channel. The unbalanced treatment of dynamic electron correlation is likely the reason for this divergence, as discussed in Ref. [64] in another context.

All other methods predict much shorter $S_2$ lifetimes. TDDFT and ADC(2) also diverge from MCSCF, predicting $S_2/S_1$ internal conversion via the ring-closed mechanism. The $S_1/S_0$ internal





conversion was consensually predicted to be the ring-open mechanism. TDDFT and MRCI/ODM3 also diverge from MCSCF regarding the photoproduct. Both predict significant C2 product formation. Methodological limitations turned ADC(2) and CC2 inadequate for these simulations.

Compared to the experimental results from Ref.[23], TDDFT-TBDA dynamics performed the best. However, such comparison defeats the purpose of the Prediction Challenge of making predictions on unseen empirical data. If the experimental information had not been available, we would not been able to pick the best prediction among those in Table 5.

Table 5. Summary of prediction from surface hopping with different electronic structure methods. Relevant secondary channels are not mentioned to be concise. For details, see Table 3, Figure 6, Figure 7, and Table 4.

| Method | $S_2$ lifetime (ps) | Dominant mechanism | | Dominant photoproducts |
|---|---|---|---|---|
| DC-FSSH with | | $S_2/S_1$ | $S_1/S_0$ | |
| MCSCF-NACV | 9.0 | Open ring | Open ring | C3 products |
| MCSCF-TDBA | 10.4 | Open ring | Open ring | C3 products |
| TDDFT-TDBA | 0.4 | Closed ring | Open ring | C2 products |
| MRCI/ODM3-NACV | 0.04 | Open ring | Open ring | C2 products |
| ADC(2)-TDBA | 0.9 | Closed ring | - | - |
| CC2-TDBA | - | - | - | - |

We commend the Journal of Chemical Physics for implementing this Prediction Challenge initiative. Metanalyses in psychology, ecology, and evolutionary biology have demonstrated that when different research groups are invited to analyze the same statistical data set, they may arrive at radically different conclusions.[65–67] We see no reason that would differ in our field, maybe even more in the Prediction Challenge, considering that each group will use different methodologies, software, and analysis protocols. At the time of writing, we still do not know how different the predictions of the diverse groups answering this challenge will be. However, based only on the striking differences in our simulations following the same protocols, we expect they will fit on a wide range.

Nonadiabatic dynamics simulations have proved invaluable for helping assign experimental features, rationalize empirical data, and explore the configurational space to reveal excited state reaction mechanisms far from our chemical intuition. Nevertheless, these conflicting dynamics results we report here (and others have reported before; for instance, Refs. [11,12]) draw a worrisome picture of a field where





routine nonadiabatic dynamics methods may have low prediction power. We hope the debate on these divergent results will lead us to more accurate and precise methodologies.

The philosopher of science, Imre Lakatos, proposed we should judge a research program as progressive if it predicted novel empirical facts and at least some novel facts could be tested.[68] If not, the program is degenerating. Nonadiabatic dynamics constitutes a research program in Lakatos' sense of a core set of theoretical assumptions (the Born-Oppenheimer separability, for example) surrounded by a protective belt of auxiliary hypotheses (the need for nonadiabatic corrections). Therefore, the utmost importance of this Prediction Challenge is to unveil the best practices and development directions we must follow to guarantee that our research field remains progressive.

## Supplementary Material

The supplementary material contains molecular orbitals at the $S_0$ minimum, optimized geometric parameters, conical intersections characterization, $S_1$-band absorption spectrum, ADC(2) dynamics results, additional results from MCSCF dynamics, working equations to simulate GUED patterns, and GUED signals for initial conditions.

## Acknowledgments

SM, RSM, JMT, and MB thank the European Research Council (ERC) Advanced grant SubNano (grant agreement 832237). The authors acknowledge the Centre de Calcul Intensif d'Aix-Marseille for granting access to its high-performance computing resources. This work was granted access to the HPC resources of TGCC under the allocation *2023-AD010813035R2* made by GENCI.

## Author Declarations

### Conflict of Interest

MB is an advisory board member of J. Chem. Phys. and guest editor of this issue. The other authors have no conflicts to disclose.

### Author Contributions

Supervision, Project Administration, Funding Acquisition: MB; Methodology: MB, HL; Analysis, Investigation: SM, RSM, JMT; Visualization: SM, RSM; Writing – Original Draft: SM, MB; Writing – Review & Editing: SM, RSM, HL, JMT, MB.





## Data availability

The data supporting this study's findings are openly available in Zenodo (DOI: 10.5281/zenodo.10662306).

# Prediction Challenge: Simulating Rydberg Photoexcited Cyclobutanone with Surface Hopping Dynamics based on Different Electronic Structure Methods


Saikat Mukherjee,[1]* Rafael S. Mattos,[2] Josene M. Toldo,[3] Hans Lischka,[4] Mario Barbatti[5,6]*

*[1]Aix Marseille University, CNRS, ICR, Marseille 13397, France; ORCID: 0000-0002-0025-4735*
*[2]Aix Marseille University, CNRS, ICR, Marseille 13397, France; ORCID: 0000-0003-0215-7100*
*[3]Aix Marseille University, CNRS, ICR, Marseille 13397, France; ORCID: 0000-0002-8969-6635*
*[4]Department of Chemistry and Biochemistry, Texas Tech University, Lubbock, TX 79409-1061 USA; ORCID: 0000-0002-5656-3975*
*[5]Aix Marseille University, CNRS, ICR, Marseille 13397, France; ORCID: 0000-0001-9336-6607*
*[6]Institut Universitaire de France, Paris 75231, France*

*\* Corresponding author: saikat.mukherjee@univ-amu.fr, mario.barbatti@univ-amu.fr*


## Supplementary Material

## Table of Contents







## S1  Molecular orbitals at S$_0$ minimum

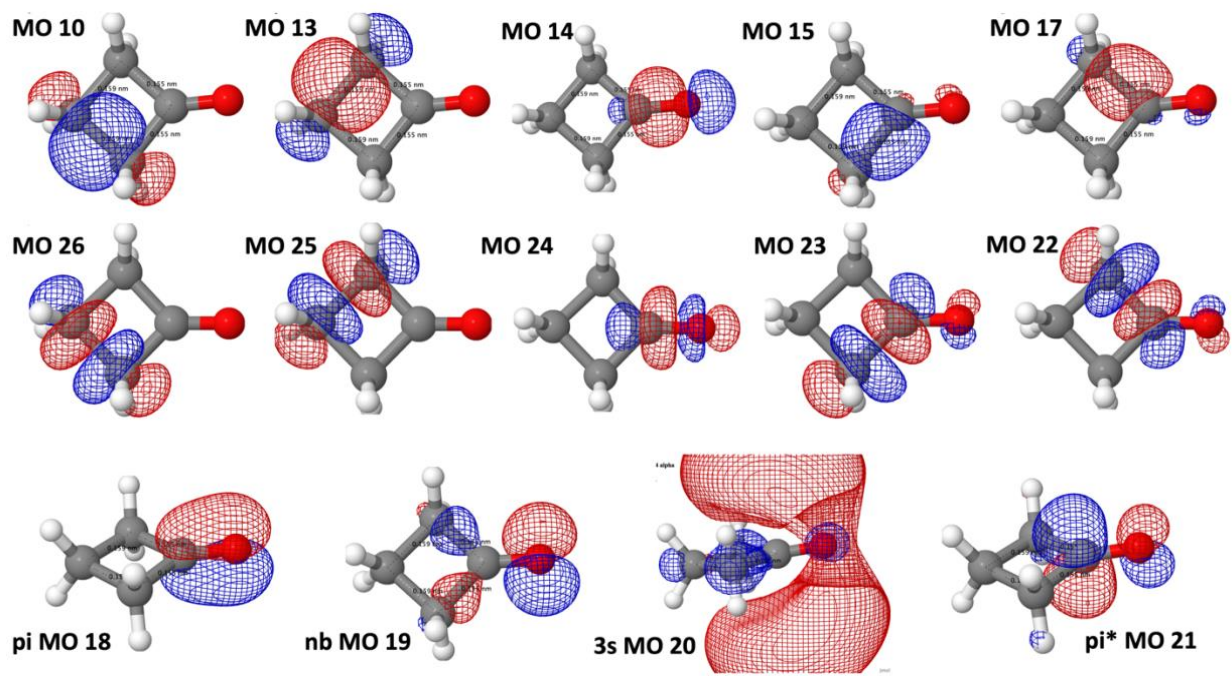

Figure S1. Molecular orbitals (MO) used in the SA(3)-MCSCF(14,14) calculation (4[PP$_{CC}$] × [PP$_{CO}$] × [CAS(4,4)$_{n,\pi,\pi^*,3s}$]). The first row shows five occupied σ orbitals; their counterpart vacant σ* orbitals are shown in the second row. The perfect-pairing (PP) multiconfigurational subspace comprises one occupied σ and one vacant σ* orbitals. In this case, a PP$_{C-C}$ contains the MO numbers 10 and 26, whereas the PP$_{C-O}$ contains the MO numbers 14 and 24. The CAS(4,4) includes MOs 18-21, that is the $\pi$, $\pi^*$ and the nonbonding and 3s orbital of oxygen, shown in the third row.





## S2 Optimized geometries

Table S1. Selected bond lengths and dihedral angles obtained at the optimized structures of cyclobutanone in the $S_0$, $S_1$, and $S_2$ minima at the MCSCF level.. The numbers in parentheses were obtained from the (TD)DFT calculations; those in square brackets were obtained from the MRCI/ODM3 calculations.

| | $S_0$ min | $S_1$ min | $S_2$ min |
|---|---|---|---|
| α C-C (Å) | 1.542 | 1.531 | 1.516 |
| | (1.529) | | |
| | [1.536] | | |
| | 1.572[#] | 1.565[#] | |
| | 1.524[##] | 1.521[##] | |
| | 1.533[$] | 1.523[$], 1.570[$] | |
| | 1.53[$$] | 1.55[$$] | 1.60[$$] |
| β C-C (Å) | 1.585 | 1.583 | 1.586 |
| | (1.554) | | |
| | [1.545] | | |
| | 1.585[#] | 1.581[#] | |
| | 1.587[##] | 1.551[##] | |
| | 1.553[$] | 1.548[$], 1.547[$] | |
| C=O (Å) | 1.215 | 1.387 | 1.236 |
| | (1.200) | | |
| | [1.219] | | |
| | 1.19[#] | 1.328[#] | |
| | 1.20[##] | 1.373[##] | |
| | 1.21[$] | 1.354[$] | |
| | 1.20[$$] | 1.28[$$] | 1.16[$$] |
| Dihedral angle C-C-C-C (°) | 2.5 | 14.4 | 0.0 |
| | (3.8) | | |
| | [0.0] | | |
| Dihedral angle O-C-C-C (°) | 176.7 | 146.0 | 0.0 |
| | (174.6) | | |
| | [180.0] | | |
| | 177.3[$$] | 144.5[$$] | 180.0[$$] |

[#]SA(2)-CAS(12,11)/6-31G*;[1] [##]CASSCF(10,8)/6-31G(d);[2] [$]CASSCF(10,8)/6-31+G*;[3] [$$]EOM-CCSD/cc-pVTZ+1s1p1d.[4]





## S3  Conical intersections characterization

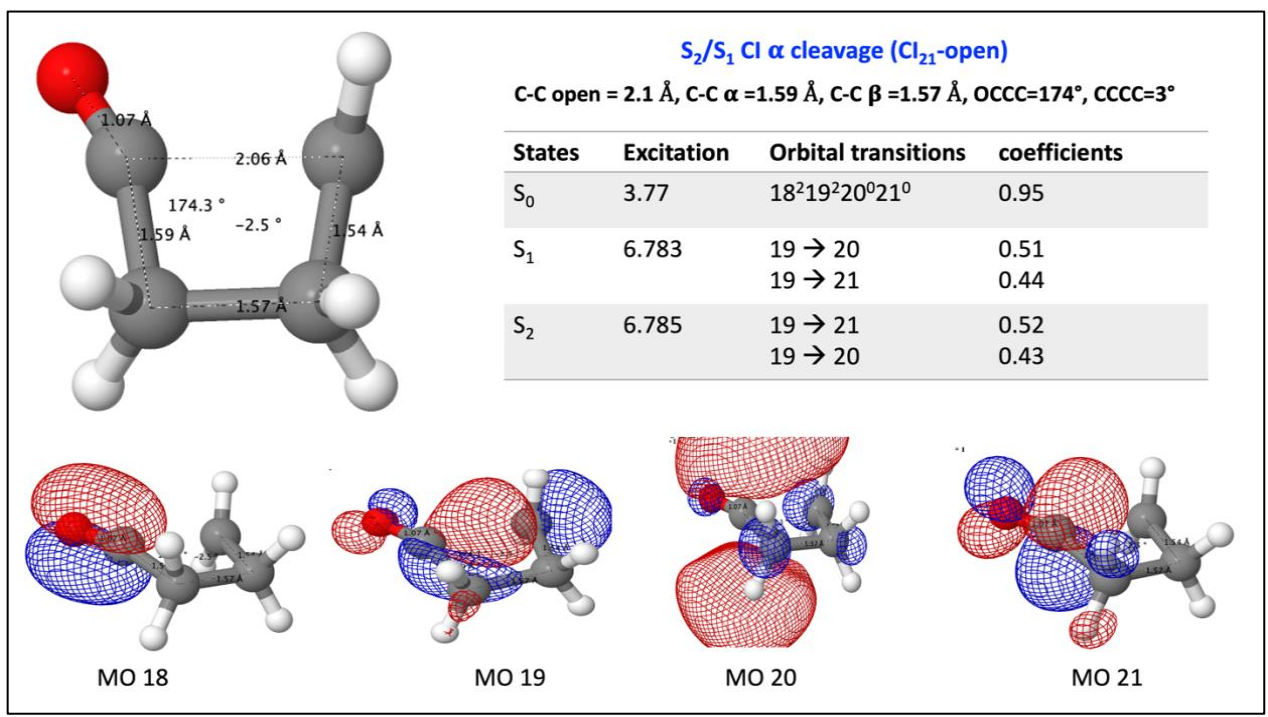

Figure S2.       Excitation energies, orbital transitions and selected geometrical features for CI₂₁-open.

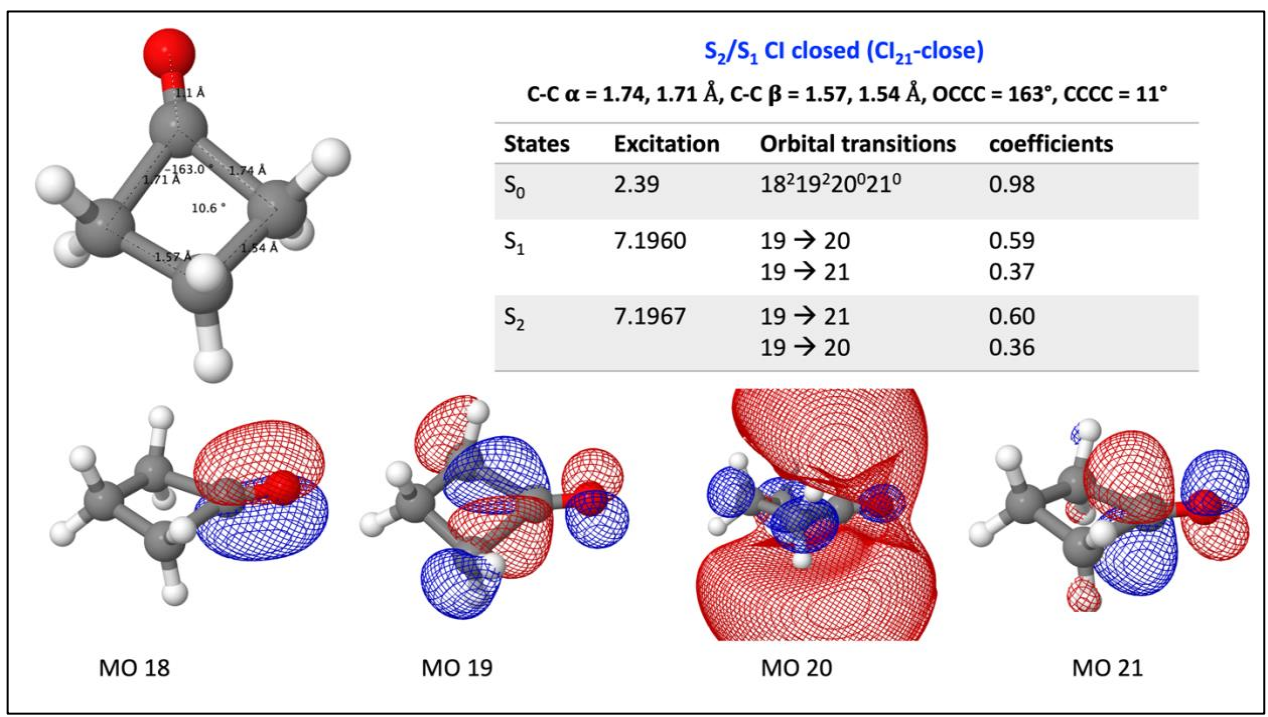

Figure S3.       Excitation energies, orbital transitions and geometrical features for CI₂₁-closed.





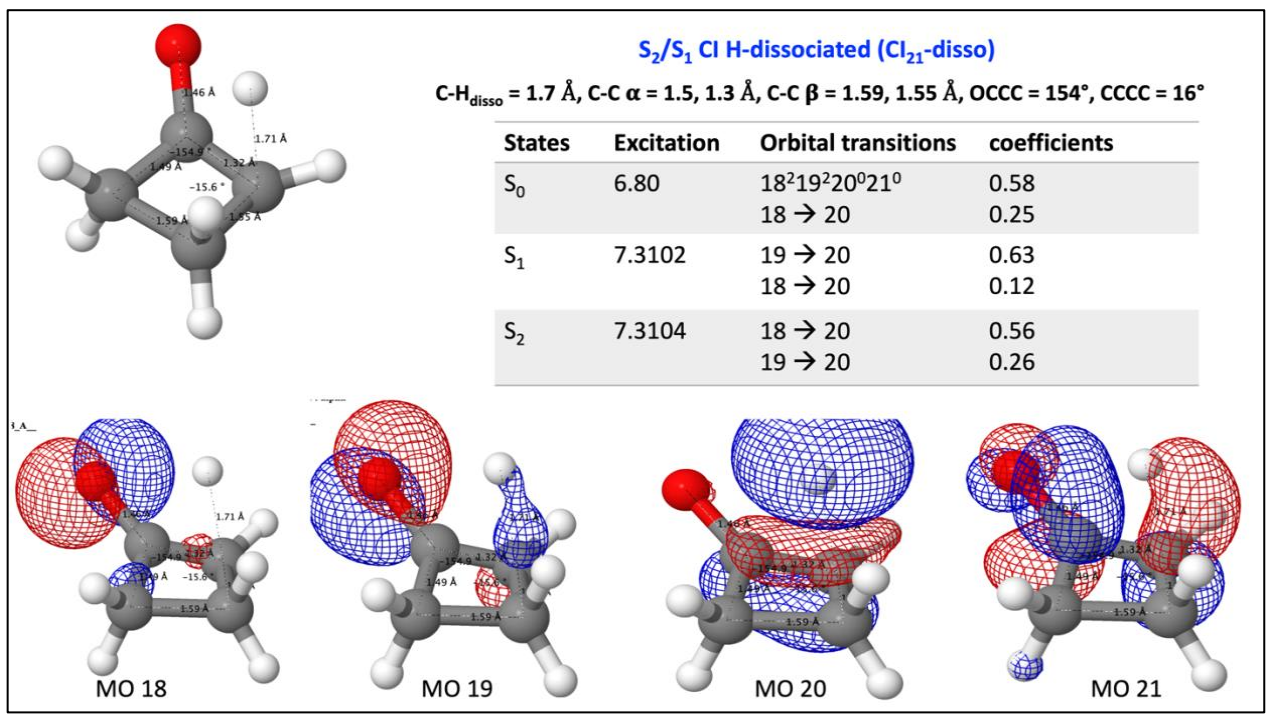

Figure S4.     Excitation energies, orbital transitions, and selected geometrical features for CI$_{21}$-Hdisso.

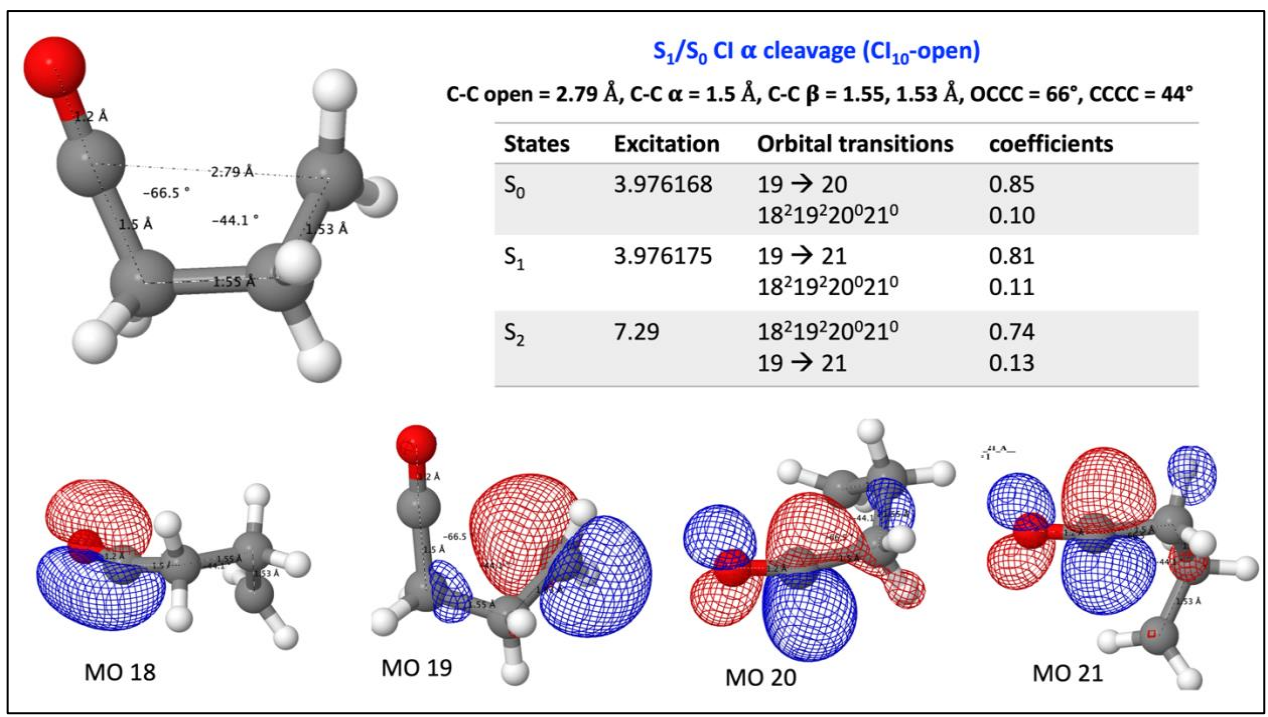

Figure S5.     Excitation energies, orbital transitions, and selected geometrical features for CI$_{10}$-open.





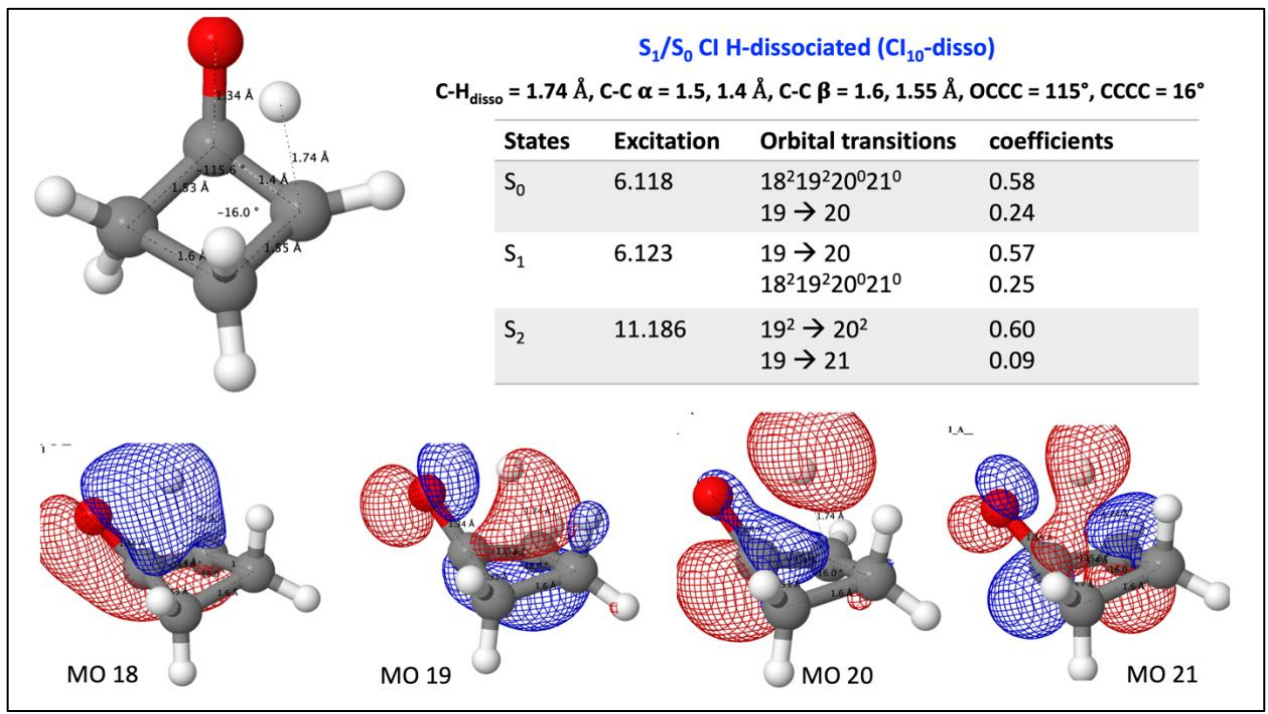

Figure S6. Excitation energies, orbital transitions, and selected geometrical features for CI$_{10}$-Hdisso.





## S4  Absorption spectra of the S$_1$ band

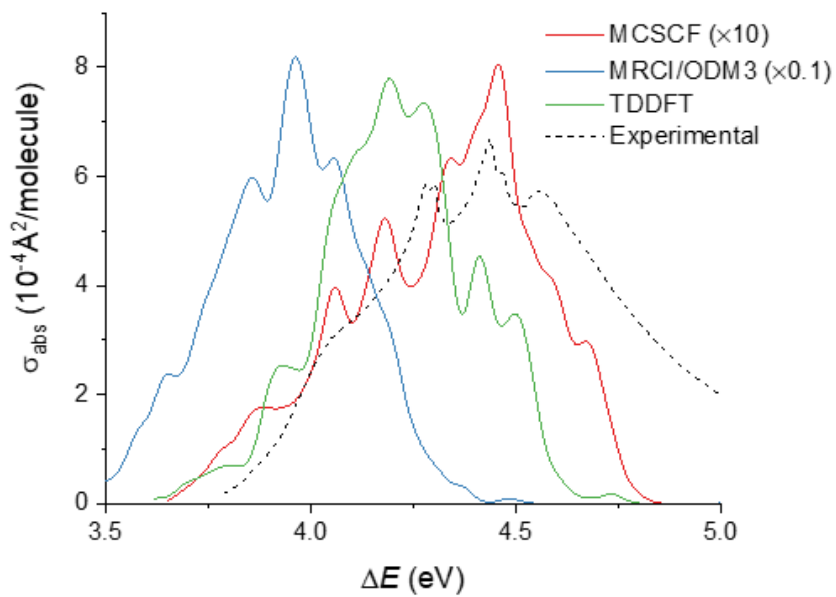

Figure S7.    Simulated and experimental[5] S$_1$ ($n\pi^\star$) absorption band of CB in the gas phase. Note the scaling factors in the MCSCF and MRCI/ODM3 simulations.





## S5  ADC(2)-TDBA nonadiabatic dynamics results

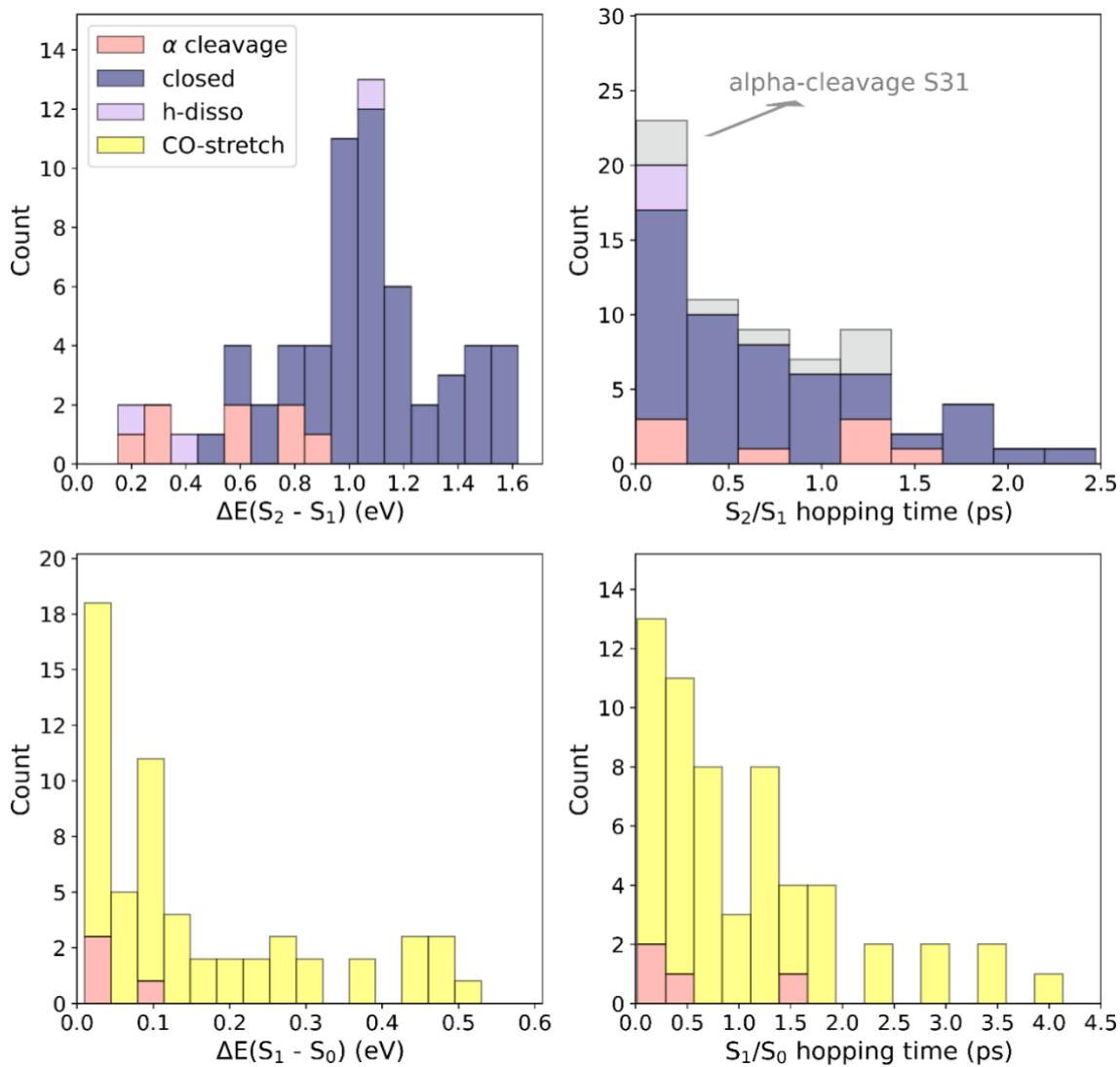

Figure S8.    Distribution of energy gaps ($\Delta E$) and hopping times at the hopping geometries using ADC(2) method. Top: S at $S_2/S_1$ hopping geometries. Bottom: at $S_1/S_0$ hopping geometries.





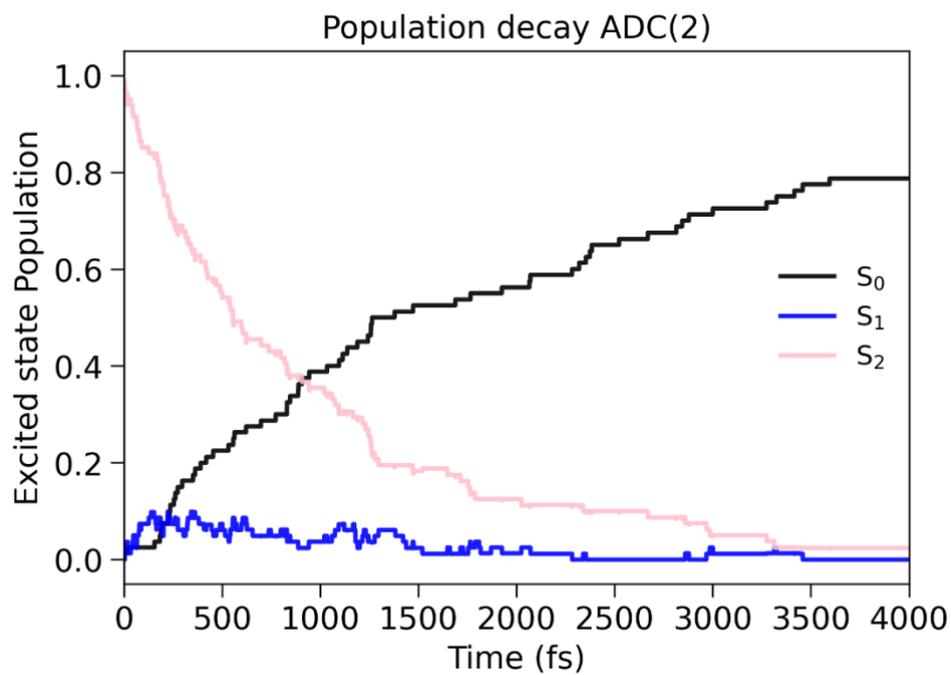

Figure S9.    Time evolution of the average fraction of excited state population of the three lowest singlet states using ADC(2) method. The fitting of the $S_2$ occupation decay provided a lifetime of 0.9 ps.





## S6 Additional results from MCSCF-NACV dynamics

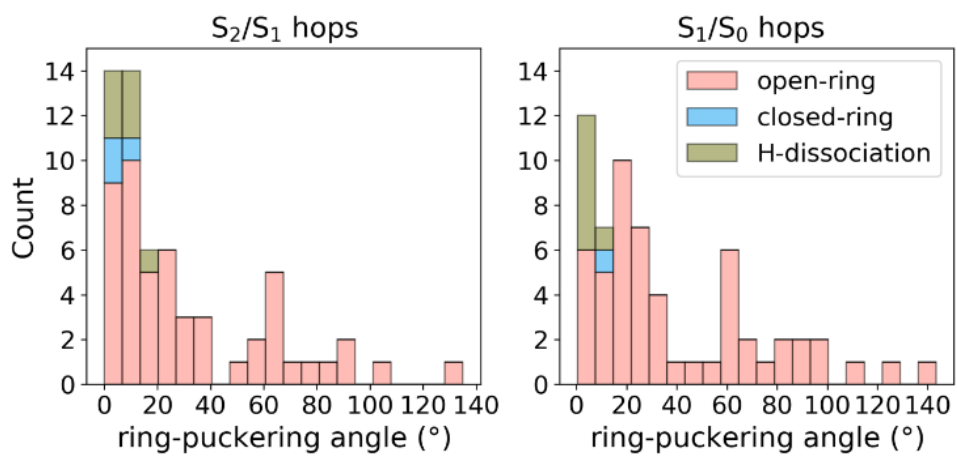

Figure S10.    Distribution of the C-C-C-C dihedral angles (ring-puckering angle) for $S_2/S_1$ and $S_1/S_0$ hopping geometries.

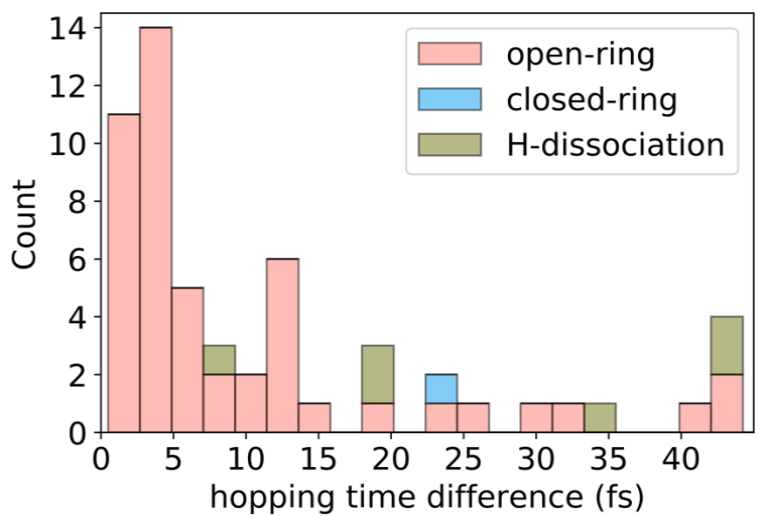

Figure S11.    The distribution of time difference between $S_2/S_1$ and $S_1/S_0$ hoppings.





## S7  Working equations to simulate GUED patterns

We simulated the GUED signal using the independent atom model (IAM) as prescribed by Centurion et al.[6] This model assumes that the electron arrangement within a molecule mimics that of independent atoms with no interactions among them. As a result, we can safely work with the predetermined atomic force factors (AFFs) to describe the arrangements. However, the approximation overlooks the alterations in electron distribution resulting from the formation of chemical bonds.

The total elastic scattering intensity, $I(s)$, of a molecule as a function of momentum transfer of the scattered electron, $s$, is a contribution of individual (atomic) and pairwise (pair) scattering amplitudes

$$I(s) = I_{atom}(s) + I_{pair}(s).$$

The atomic scattering contribution, $I_{atom}(s)$, is the sum of each atomic differential cross-section

$$I_{atom}(s) = \sum_{i=1}^{N} |f_i(s)|^2,$$

where the AFFs, $f(s)$, is given by ELSEPA program[7,8] and the sum runs over all $N$ atoms.

The pair-wise scattering contribution is calculated for each possible atom pairs defined by the internuclear distances, $r_{ij} = \|\vec{r}_i - \vec{r}_j\|$ between atoms $i$ and $j$.

$$I_{pair}(s) = \sum_{i=1}^{N} \sum_{j \neq i}^{N} |f_i(s)||f_j(s)| \frac{\sin(sr_{ij})}{sr_{ij}}.$$

Hence, the overall elastic scattering signal is the incoherent sum over an ensemble of all the molecular structures obtained form the nonadiabatic dynamics simulation. It is customary to work with the modified scattering intensity,

$$sM(s) = s \frac{I_{pair}(s)}{I_{atom}(s)}.$$

The real-space pair distribution function, $PDF(r)$, as a function of atom-pair distance, is,

$$PDF(r) = r \int_0^\infty sM(s) \sin(sr)\, ds.$$

In practice, the limits of the integration consider the experimental condition, such as the limits of the detector, $s_{min}$ to $s_{max}$ and adding a damping factor that avoids sharp-edged sine-transform effects at the high $s$ values, which acts as a Gaussian smoothing in real space.

$$PDF(r) = r \int_{s_{min}}^{s_{max}} sM(s) \sin(sr)\, e^{-\alpha s^2} ds.$$





The time-dependent signals, $\Delta PDF(r,t)$, are obtained by subtracting the modified scattering intensity at time zero as,

$$\Delta PDF(r,t) = r \int_0^{s_{max}} [sM(s,t) - sM(s,t=0)] \sin(sr) \, e^{-\alpha s^2} ds,$$

where the signal of $sM(s,t=0)$ is the average quantity of all trajectories at time $t=0$.

In this work, we used "Diffraction_simulation" scripts developed by Wolf et al. (github.com/ThomasJAWolf/Diffraction_simulation) to simulate the time-dependent modified scattering intensity from molecular geometries obtained from nonadiabatic dynamics simulations. The repository contains a file "Xsects.mat" containing angle-dependent scattering cross-sections ($f(s)$) for all elements of the periodic table, which were simulated using the ELSEPA program (github.com/eScatter/elsepa) for different elements.

## S8 GUED signal for the initial conditions

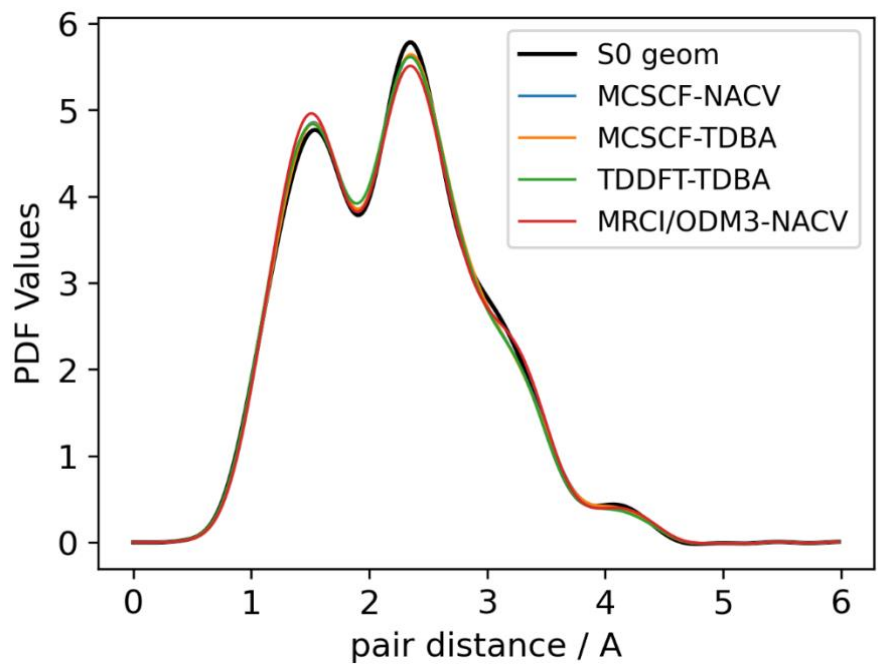

Figure 12. Pair distribution function for the ground state computed with MCSCF and the average values for the initial conditions used for the different sets of simulations.

## S9    Supplementary References